\documentclass[sigconf]{acmart}
\usepackage{multirow}

\AtBeginDocument{%
  \providecommand\BibTeX{{%
    \normalfont B\kern-0.5em{\scshape i\kern-0.25em b}\kern-0.8em\TeX}}}

\setcopyright{acmcopyright}
\copyrightyear{2022}
\acmYear{2022}
\acmDOI{XXXXXXX.XXXXXXX}

\acmConference[KDD '22]{The 28th ACM SIGKDD Conference on Knowledge Discovery and Data Mining}{August 14--18,
  2022}{Washington, DC}

\acmPrice{15.00}
\acmISBN{978-1-4503-XXXX-X/18/06}

\begin{document}

\title{ItemSage: Learning Product Embeddings for Shopping Recommendations at Pinterest}

\author{Paul Baltescu}
\email{pauldb@pinterest.com}
\affiliation{%
  \institution{Pinterest}
  \city{San Francisco}
  \country{USA}
}

\author{Haoyu Chen}
\email{hchen@pinterest.com}
\affiliation{%
  \institution{Pinterest}
  \city{San Francisco}
  \country{USA}
}

\author{Nikil Pancha}
\email{npancha@pinterest.com}
\affiliation{%
  \institution{Pinterest}
  \city{San Francisco}
  \country{USA}
}

\author{Andrew Zhai}
\email{andrew@pinterest.com}
\affiliation{%
  \institution{Pinterest}
  \city{San Francisco}
  \country{USA}
}

\author{Jure Leskovec}
\email{jure@pinterest.com}
\affiliation{%
  \institution{Pinterest}
  \city{San Francisco}
  \country{USA}
}

\author{Charles Rosenberg}
\email{crosenberg@pinterest.com}
\affiliation{%
  \institution{Pinterest}
  \city{San Francisco}
  \country{USA}
}

\begin{abstract}
Learned embeddings for products are an important building block for web-scale e-commerce recommendation systems. At Pinterest, we build a single set of product embeddings called ItemSage to provide relevant recommendations in all shopping use cases including user, image and search based recommendations. This approach has led to significant improvements in engagement and conversion metrics, while reducing both infrastructure and maintenance cost. While most prior work focuses on building product embeddings from features coming from a single modality, we introduce a transformer-based architecture capable of aggregating information from both text and image modalities and show that it significantly outperforms single modality baselines. We also utilize multi-task learning to make ItemSage optimized for several engagement types, leading to a candidate generation system that is efficient for all of the engagement objectives of the end-to-end recommendation system. Extensive offline experiments are conducted to illustrate the effectiveness of our approach and results from online A/B experiments show substantial gains in key business metrics (up to +7\% gross merchandise value/user and +11\% click volume).
\end{abstract}

\ccsdesc[500]{Information systems~Recommender systems}
\ccsdesc[500]{Information systems~Multimedia and multimodal retrieval}
\ccsdesc[500]{Computing methodologies~Multi-task learning}

\keywords{Representation Learning, Multi-Task Learning, Multi-Modal Learning, Recommender Systems}


\maketitle

\section{Introduction}
\begin{figure}
    \centering
    \includegraphics[width=\columnwidth]{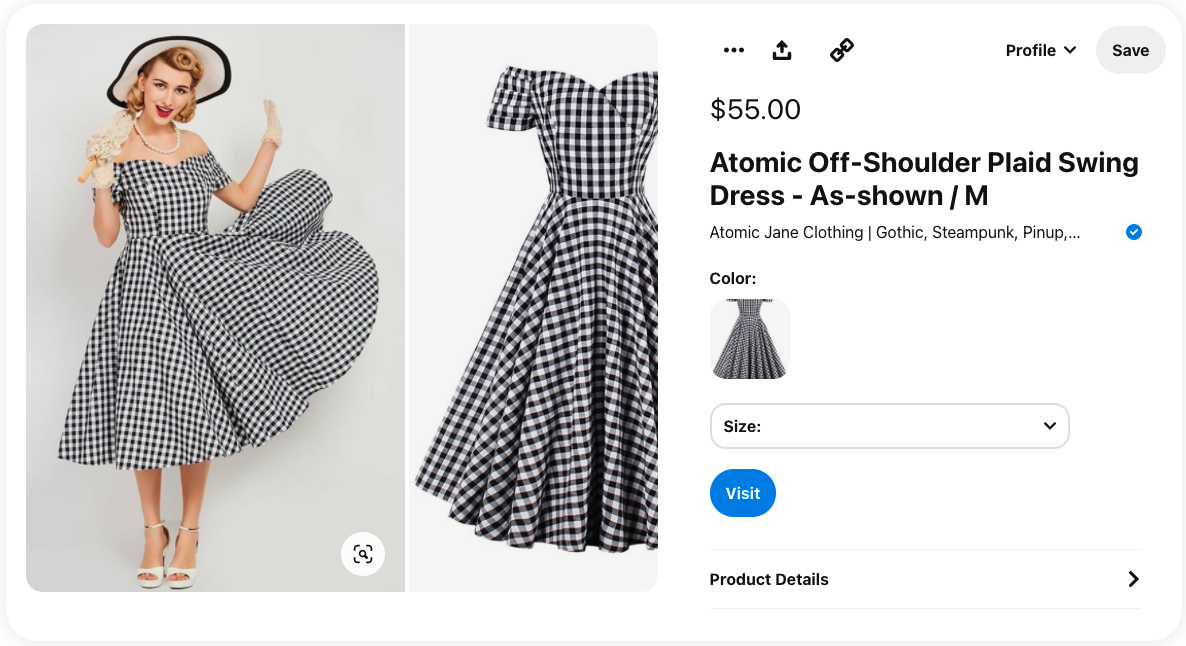}
    \caption{Products on Pinterest consist of several images and rich textual metadata.}
    \label{fig:product_item}
\end{figure}

Pinterest’s mission is to bring everyone the inspiration to create a life they love. Users browse Pinterest to get inspired for their next home decoration project or to stay up to date with the latest fashion and beauty trends. Common feedback we hear from our users is that once they discover a product that matches their taste, they want to be able to purchase it as seamlessly as possible. In order to build a delightful shopping experience, we need our recommendation systems to evolve beyond image-based recommendations by leveraging the multi-faceted information available for products in our shopping catalog. Unlike pins - the main type of content on Pinterest, products consist of several images displaying the product from different angles or in different contexts and have high quality textual metadata provided by merchants including title, description, colors, and patterns in which the product is available for sale (see Figure \ref{fig:product_item} for an example). Our shopping recommendation systems also need to optimize for new types of outcomes like purchases and add-to-cart actions in addition to typical engagement metrics like clicks or saves.

This paper introduces Pinterest’s learned embedding representation for products named ItemSage. Embeddings are a powerful tool for building recommendation systems at scale for e-commerce \cite{amazonsearch,amazonsearch2,jd} and social media  \citep{pinsage,visual,pinnersage,youtube,facebooksearch,twitter} applications. From our experience, one of the key reasons to focus on building embedding representations lies in their versatility: we have successfully experimented with using ItemSage embeddings (1) for generating candidates to upstream ranking systems via approximate nearest neighbor search, (2) as features in the ranking models responsible for determining the final ordering of the products shown to users and (3) as signals in classification models aimed at inferring missing information from the shopping catalog (e.g. the category or the gender affinity for a specific product).

With ItemSage, we made a few conscious design decisions that contrast it from other approaches in several key aspects:

\textbf{Multi-modal features.} Earlier approaches typically focus on building embeddings from content in a single modality, e.g. text \cite{pintext,amazonsearch,detext} or images \cite{visual,visualtransformer}. Product information spans both modalities. Since Pinterest is a visually dominant platform, it is important to capture the nuanced information available in a product’s images to make sure shopping recommendations feel natural with the rest of the product (e.g. users tend to prefer fashion products shown in lifestyle photos over images of the products on a white background). At the same time, product images may contain other products (e.g. a sofa might be shown in a living room with a coffee table and a rug) so textual matches are crucial for providing highly relevant recommendations. We introduce a transformer-based architecture capable of combining features from these different modalities which is different from earlier work \cite{pinsage,youtube} that extends to multi-modal features.

\begin{figure}
    \centering
    \includegraphics[width=\columnwidth]{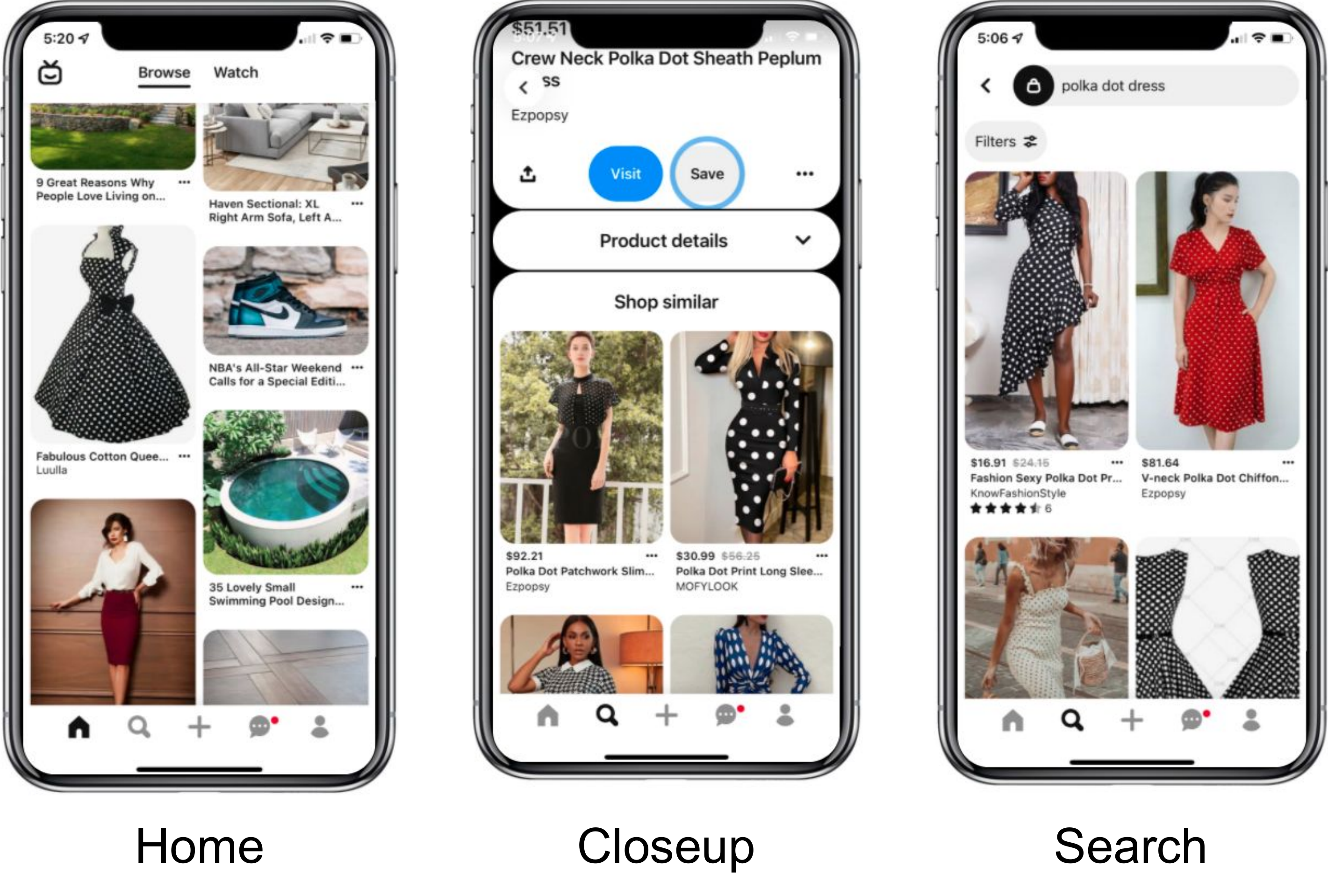}
    \caption{Screenshots of ItemSage being used for product recommendations on Home, Closeup and Search surfaces.}
    \label{fig:surfaces}
\end{figure}

\textbf{Multi-modal vertical recommendations.} Pinterest has 3 main surfaces (Figure \ref{fig:surfaces}) that provide personalized recommendations: (1) in the \textbf{Home} surface, users are provided with recommendations based on their past activity, (2) in the \textbf{Closeup} surface, we provide similar recommendations to a pin the user is currently viewing, while (3) in the \textbf{Search} surface, we provide recommendations in response to a query string that the user has typed. Note that in each surface, the query comes from a different modality: (1) in Home, the query is essentially a \textbf{sequence of pins}, (2) for Closeup, the query is a \textbf{single pin}, (3) while in Search, the query is a \textbf{text string}. In contrast with other works that typically target a single vertical application (e.g. product search \cite{amazonsearch, jd}), ItemSage can provide relevant candidates via approximate neighbor search \cite{hsnw} for all these surfaces and, therefore, in response to queries formulated in each of these modalities. We achieve this by training ItemSage embeddings to be compatible with the learned representations for pins \cite{pinsage} and search queries. Recommendations based on user activities are a more general case of pin-based recommendations where the activity history is first partitioned into clusters and then a few representative pins are sampled from different clusters to generate pin to product recommendations \cite{pinnersage}.


\textbf{Multi-task Learning.} To strike the right balance between inspiration and conversion optimization, Pinterest shopping recommendation systems optimize for several objectives including purchases, add-to-cart actions, saves and clicks. Traditionally, recommendation systems tackle multi-objective optimization in the ranking phase by building models capable of computing a score for every engagement type under consideration conditioned on the query context and the ranked candidate \cite{youtubemtl,ple}. In our work, we learn embeddings that \textbf{additionally} optimize the candidate generation phase for all engagement types under consideration, leading to an overall more efficient recommendation funnel. We show that in cases where the labels are sparse, multi-task learning leads to improved results over models specialized only on sparse objectives. For objectives where labeled data is plentiful, we show that we can optimize our embeddings for new objectives with little or no performance penalty on existing objectives.

\section{Related Work}
The main concepts that underpin our work building product embeddings at Pinterest are multi-modal learning and multi-task learning. ItemSage aggregates multi-modal information from images and text. The embeddings are learned in a multi-task regime that supports cross-modal candidate retrieval and joint optimization for several engagement types. In this section, we briefly review these two concepts and related work. Another goal of our work is to create embeddings that are \textbf{compatible} with the learned representations of other entities in the Pinterest ecosystem, namely pins and search queries. We briefly review our approach for generating these “target” embeddings.

\subsection{Multi-Modal Representation Learning}
Multi-modal representation learning aims to aggregate information from different modalities into a common subspace \cite{multimodal}. It has been extensively studied in areas like video classification \cite{videoclassification1,videoclassification2}, speech recognition \cite{speechrecognition1,speechrecognition2}, and visual question answering \cite{visualqa,tan2019lxmert} where information are often available in audio, text, and visual formats. 
The fusion of different modalities often follows a projection and concatenation pattern. For example, \cite{liu2016multimodal} first projects the image, audio, and text features into the same space using autoencoders and then concatenates the hidden features to produce the final embedding with a further projection. Inspired by the success of Transformer models like BERT \cite{bert}, more works adopt Transformer models for modality fusion \cite{tan2019lxmert,zhou2020unified}. Among these, the single-stream Transformer model \cite{chen2019uniter,li2019visualbert}, which also uses the same projection and concatenation idea, is the most suitable for our use case given our modalities.

We should note that although multi-modal representation learning has been studied in various areas, few works \cite{youtube,pinsage} 
have successfully applied it to large-scale recommendation systems. To our knowledge, this work is the first one that uses the Transformer architecture to aggregate image and text modalities to learn product representations for production-scale recommendation systems.

\subsection{Multi-Task Representation Learning}
Multi-task learning is designed to improve the model performance on individual tasks by sharing model parameters across related tasks \cite{mtloverview}. Typical multi-task learning models are deployed in the ranking phase of the recommendation systems \cite{youtubemtl,ple}, and a model will have several outputs, one for each task. Similar model architectures are used in representation learning \cite{liu2015representation} where multiple task-specific embeddings are learned by a model. However, from the production point of view, it is most convenient and economic to use a single version of embeddings for all tasks. Therefore, we take the same approach as \cite{visual,visualtransformer} and utilize multi-task learning to optimize for a single embedding. While these works are optimized for multiple classification tasks, our work is trained on retrieval tasks with several engagement types for multi-modal vertical recommendation systems. This kind of multi-modal multi-task learning helps us solve the special challenge of making our learned product embeddings compatible with embeddings of both query images and search queries.

\subsection{Image and Search Query Representations at Pinterest}\label{sec:xsage}
PinSage \cite{pinsage} is a highly-scalable implementation of the GraphSage GCN algorithm \cite{graphsage} that is deployed in production at Pinterest to produce the image embeddings of billions of pins. It aggregates the visual and textual features along with the graph information of pins to produce a rich and compact representation for various use cases including retrieval, ranking, and classification. 

SearchSage\footnote{\url{https://medium.com/pinterest-engineering/searchsage-learning-search-query-representations-at-pinterest-654f2bb887fc}} is our search query embedding model trained by fine-tuning DistilBERT \cite{distilbert}. It is trained on search query and engaged pin pairs from search logs. The loss aims to optimize the cosine similarity between the embedding of the global \texttt{[CLS]} token which can be seen as an aggregation of the input query and the output of an MLP that summarizes the pin into an embedding based on several text features and its PinSage embedding. Because other features in addition to PinSage contribute to the candidate pin representation and because of the MLP transformation, PinSage and SearchSage embeddings are not directly compatible with one another. We will refer to this later when we discuss baselines for ItemSage embeddings.

When training ItemSage, the PinSage model is used to provide the embeddings of both the feature images of the product and the query images, while the SearchSage model embeds the search query string. Since the PinSage and SearchSage models both have multiple applications in production, they are frozen at ItemSage training time due to considerations of development velocity and ease of adoption.

\section{Method}
In this section, we introduce our approach to building product embeddings at Pinterest. We first formally define the notion of compatibility across learned representations for different entities. Then, we introduce our model starting with the features used as input by the model and the process for collecting the training labels for the different tasks. We provide a detailed description of the model architecture, loss function, the multi-task learning regime and the inference and serving setup.

\subsection{Embedding Compatibility}
One of the requirements for ItemSage is to create product embeddings that are compatible with PinSage embeddings for images and SearchSage embeddings for search queries. In this case, compatibility means that the distance between a query (image or text string) embedding and a product embedding should be an informative score indicating how relevant the product is as a result for the query. We use cosine similarity as a measure of the embedding distance due to its simplicity and efficiency. The compatibility requirement originates from our desire to support candidate generation via approximate nearest neighbor (ANN) search techniques like Hierarchical Navigable Small Worlds (HNSW) \cite{hsnw}. We cannot afford to apply expensive transformations to achieve compatibility as they would significantly increase retrieval latency. On the other hand, our experience indicates that for ranking and classification applications, compatibility plays a less important role as most deep learning models operating on pretrained embeddings can learn MLP transformations that are sufficient to map embeddings into a shared latent space.

\subsection{Features and Labels}

\begin{table}
  \caption{Training data volume.}
  \label{tab:data-volume}
  \begin{tabular}{c c c c}
    \toprule
    Surface & Engagement & No. Examples \\
    \midrule
Closeup & Clicks + Saves & 93.3M \\
Closeup & Checkouts + Add-to-Cart & 3.7M \\
Search & Clicks + Saves & 99.4M \\
Search & Checkouts + Add-to-Cart & 3.5M \\
    \bottomrule
\end{tabular}
\end{table}

A product consists of a list of images depicting the product from different angles or in different contexts and a list of text features. We truncate the list of images to at most 20 (99.7\% of products in our catalog have less than or equal to 20 images). Each image is represented by its pretrained PinSage embedding \cite{pinsage} and for products with fewer than 20 images, each empty slot is represented by a zero embedding. We use 12 text features as input to our model: title, description, merchant domain, product links, google product category\footnote{The google product category represents the node from a standard taxonomy that merchants may tag their products with. The taxonomy can be found at \url{https://www.google.com/basepages/producttype/taxonomy.en-US.txt}}, product types\footnote{The product types are html breadcrumbs scraped from the merchant product page, e.g. \texttt{Home Decor > New Arrivals}.}, brand, colors, materials, patterns, size, and size type. In cases where a product may have several values for a particular feature (e.g. links, colors, etc.) these strings are concatenated into one string. Standard word-level tokenization and lowercasing are applied to all text features. Each processed string is represented as a bag of word unigrams, bigrams and character trigrams \cite{dssm, amazonsearch}. The tokens are mapped to numerical IDs using a vocabulary $\mathcal{V}$ of the most frequent 200,000 word unigrams, 1,000,000 word bigrams and 64,000 character trigrams and out-of-vocabulary tokens are discarded.

We construct our training dataset by collecting labels from the Closeup and Search engagement logs. Each positive example is a pair of query and engaged product where the query represents either an image for examples mined from Closeup logs or a text string for search logs. The dataset is deduplicated such that only one instance of a query and engaged product pair is kept. We select 4 engagement types to train our models: clicks and saves are the main actions that users can take on any Pinterest content, while add-to-cart actions and checkouts are actions that express shopping intent. The labels for all tasks are collected from the same date range. The number of positive labels is summarized in Table \ref{tab:data-volume}. In addition to positive labels, our loss uses random negative labels which we sample randomly from the shopping catalog. The negative labels are joined with the features offline and streamed into the model trainer through a separate data loader. The trainer alternates between consuming a batch of positive labels and a batch of negative labels which are then concatenated into a single training batch.

\subsection{Model Architecture}
\begin{figure}
    \centering
    \includegraphics[width=\columnwidth]{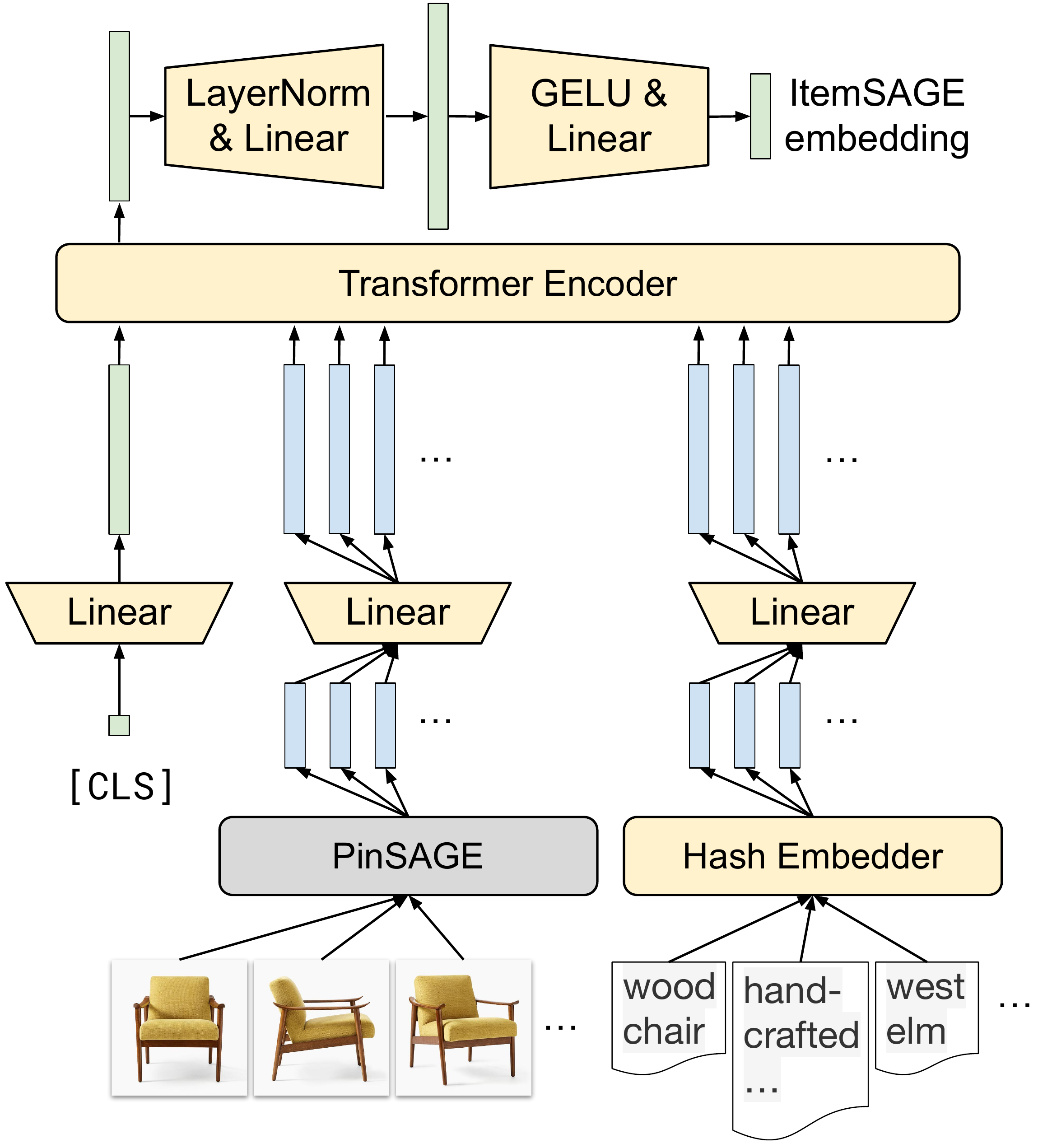}
    \caption{ItemSage model architecture}
    \label{fig:itemsage}
\end{figure}

We use a transformer encoder as the basic building block for learning product embeddings. It takes as input a sequence of 32 embeddings representing the image and text features of each product. The image embeddings are generated by the pretrained PinSage model \cite{pinsage}, while the text embeddings are learned jointly with the encoder. In order to deal with the large vocabulary size, we apply the hash embedding trick from \cite{svenstrup2017hash} to learn a compact embedding table. 
The hash embedder consists of an embedding table $E$ of size $100,000 \times 256$ and an importance weight table $W$ of size $|\mathcal{V}| \times 2$. We use two hashing functions $h_1$, $h_2$ to map each token ID $i = 1, \ldots, |\mathcal{V}|$ into two slots in the embedding table $h_1(i)$, $h_2(i)$. The embedding of token with ID $i$ is then the weighted interpolation of the two embeddings: $W_{i1} E_{h_1(i)} + W_{i2} E_{h_2(i)}$. The final embedding of a feature string is the summation of all its token embeddings.

As shown in Figure \ref{fig:itemsage}, we apply a linear transformation with output size 512 for each group of feature embeddings. This allows us to relax the requirement that all input features must have the same dimension. Similar to \cite{bert}, we use a global token \texttt{[CLS]} to aggregate the information from the input sequence. The transformed embeddings are concatenated together with the global token and then passed through a one-layer transformer block consisting of a self-attention module with 8 heads and a feed forward module with one hidden layer. The output corresponding to the global token then goes through a two-layer MLP head to produce the 256-dimension product embedding. The final ItemSage embedding is $L_2$-normalized for easier computation of cosine similarity with query embeddings when performing ANN search. We experimented with deeper transformer encoders in Section \ref{sec:arch-experiments}, but did not see an improvement in offline metrics.

\subsection{Loss} \label{sec:loss}
We frame the problem of learning product embeddings as an extreme classification problem where given a query entity, our goal is to predict the product from the catalog that the user will engage with next \cite{youtube}. Formally, let $\{x_i, y_i\}_{i=1}^{|\mathcal{B}|}$ be a training batch of query, engaged product pairs sampled from the engagement logs and $\mathcal{B} = \{y_i\}_{i=1}^{|\mathcal{B}|}$ be the set of engaged products in the batch. Let $\mathcal{C}$ denote the catalog containing all products. Given the pretrained embeddings $q_{x_i} \in \mathbb{R}^d$ for the queries $x_i$, our goal is to learn embeddings $p_{y_i} \in \mathbb{R}^d$ for the engaged products $y_i$ such that $p_{y_i}$ is more similar to $q_{x_i}$ than to all of the embeddings of other products from the catalog. This can be achieved by minimizing the softmax loss
\begin{equation}
     L_{S} = -\frac{1}{|\mathcal{B}|} \sum_{i=1}^{|\mathcal{B}|} \log \frac{ e^{\langle q_{x_i}, p_{y_i} \rangle}  }{ \sum_{y \in \mathcal{C}} e^{\langle q_{x_i}, p_y \rangle} },
\end{equation}
where $\langle \cdot, \cdot \rangle$ denotes the dot product function. In our case $\langle q_{x_i}, p_{y_i} \rangle$ is the same as the cosine similarity between the two embeddings since they are $L_2$-normalized.

The main challenge with computing the softmax loss $L_{S}$ is the expensive nature of the normalization step $\sum_{y \in \mathcal{C}} e^{\langle q_{x_i}, p_y \rangle}$ which involves a summation over the entire catalog. To make the loss computation tractable, a common technique is to approximate the normalization term by treating all the other positive examples from the same training batch as negatives and ignoring all the remaining products in the catalog. This approach is very efficient as no additional product embeddings need to be generated to compute the loss. However, naively replacing the whole catalog $\mathcal{C}$ with $\mathcal{B}$ introduces a sampling bias to the full softmax. We address this issue by applying the logQ correction \cite{corrected_softmax} that updates the logits $\langle q_{x_i}, p_{y} \rangle$ to be $\langle q_{x_i}, p_{y} \rangle - \log Q_p(y | x_i)$, where $Q_p(y | x_i)$ is the probability of $y$ being included as a positive example in the training batch. The loss becomes:
\begin{equation}\label{eq:sampled_softmax_pos}
     L_{S_{pos}} = -\frac{1}{|\mathcal{B}|} \sum_{i=1}^{|\mathcal{B}|} \log \frac{e^{\langle q_{x_i}, p_{y_i} \rangle - \log Q_p(y_i | x_i)}}{\sum_{y \in \mathcal{B}} e^{\langle q_{x_i}, p_y \rangle - \log Q_p(y | x_i)}},
\end{equation}
We estimate the probabilities $Q_p(y | x_i)$ in streaming fashion using a count-min sketch that tracks the frequencies with which entities appear in the training data. The count-min sketch \cite{cms} is a probabilistic data structure useful for tracking the frequency of events in streams of data that uses sub-linear memory.

One problem we have experienced with using in-batch positives as negatives is that unengaged products in the catalog will never appear as negative labels. This treatment unfairly penalizes popular products as they are ever more likely to be selected as negatives. To counter this effect, we adopt a mixed negative sampling approach inspired by \cite{yang2020mixed}. For each training batch, we further select a set of random negatives $\mathcal{N}$ (where $|\mathcal{N}|=|\mathcal{B}|$) based on which we compute a second loss term:
\begin{equation}\label{eq:sampled_softmax_neg}
     L_{S_{neg}} = -\frac{1}{|\mathcal{N}|} \sum_{i=1}^{|\mathcal{N}|} \log \frac{e^{\langle q_{x_i}, p_{y_i} \rangle - \log Q_n(y_i)}}{\sum_{y \in \mathcal{N}} e^{\langle q_{x_i}, p_y \rangle - \log Q_n(y)}},
\end{equation}
where $Q_n(y)$ represents the probability of random sampling product $y$. The loss term $L_{S_{neg}}$ helps reduce the negative contribution that popular products receive when used as negative labels. The main distinction between our approach and \cite{yang2020mixed} is that we optimize for $L_{S_{pos}} + L_{S_{neg}}$, while \cite{yang2020mixed} uses both $\mathcal{B}$ and $\mathcal{N}$ to calculate the normalization term and minimizes
\begin{equation}\label{eq:sampled_softmax_mixed}
\begin{split}
     L_{S_{mixed}} &= -\frac{1}{|\mathcal{B}|} \sum_{i=1}^{|\mathcal{B}|} \log \frac{e^{\langle q_{x_i}, p_{y_i} \rangle - \log Q_p(y_i | x_i)}}{Z} \\
     Z &= \sum_{y \in \mathcal{B}} e^{\langle q_{x_i}, p_y \rangle - \log Q_p(y | x_i)} + \sum_{y \in \mathcal{N}} e^{\langle q_{x_i}, p_y \rangle - \log Q_n(y)}.
\end{split}
\end{equation}
Section \ref{sec:loss-ablation} shows that we obtain better results separating the loss terms of the two negative sampling approaches.

\subsection{Multi-task Learning}

\begin{figure}
    \centering
    \includegraphics[width=\columnwidth]{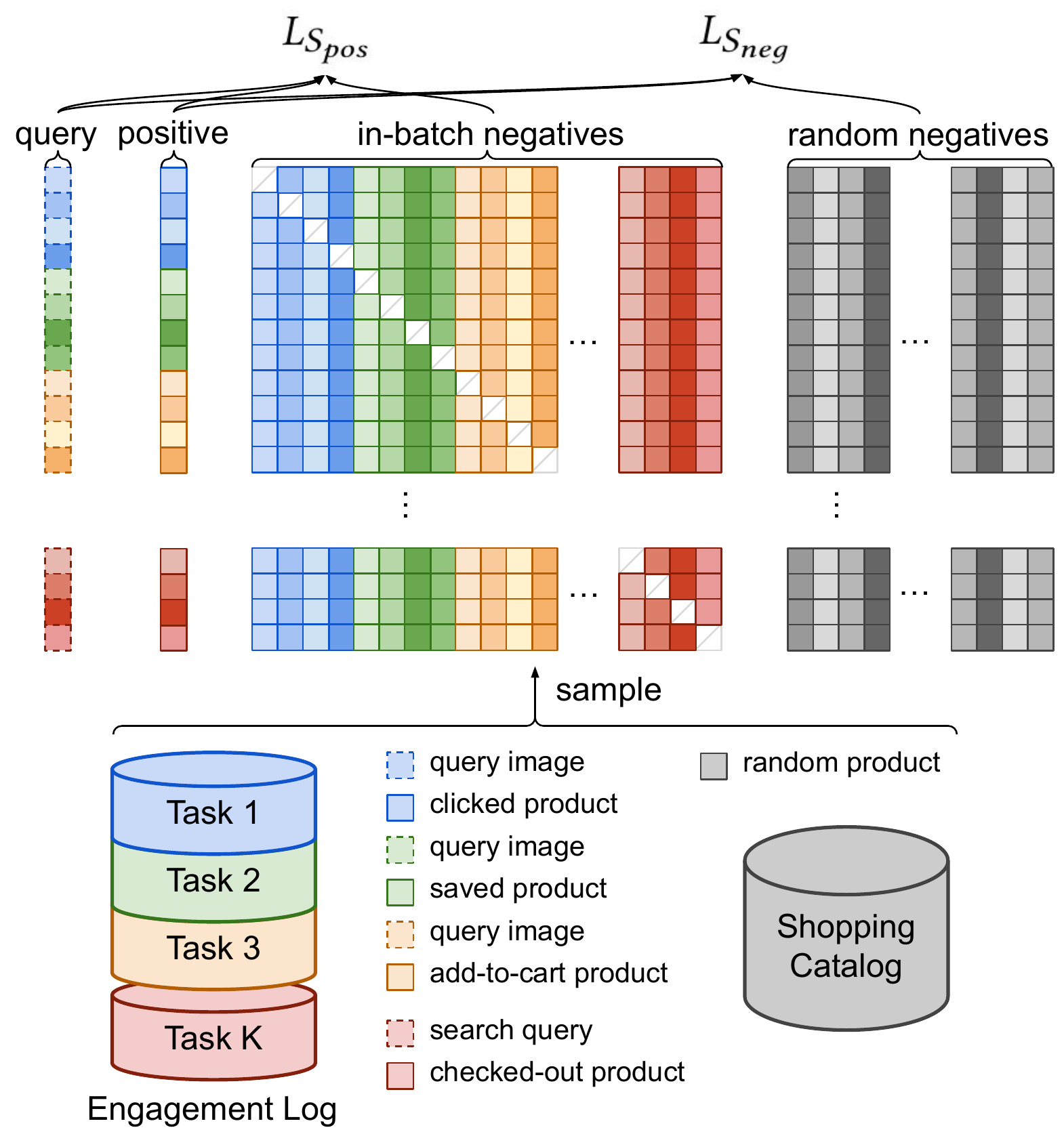}
    \caption{The construction of a training batch. The white squares on the diagonal indicates a mask applied to prevent using an example's positive label also as a negative. Squares of different shades denote different queries and products.}
    \label{fig:training_batch}
\end{figure}

We implement multi-task learning by combining positive labels from all the different tasks into the same training batch (Figure \ref{fig:training_batch}). This technique is effective even when the query entities come from different modalities, the only difference being that the query embedding needs to be inferred with a different pretrained model. In a training batch of size $|\mathcal{B}|$, we allocate $T_k$ positive examples for each task of interest $k \in \{1, \cdots, K\}$ such that $|\mathcal{B}| = \sum_{k=1}^K T_k$. Therefore tuning the values $T_k$ to be an effective way to create trade-offs between different tasks. When introducing new tasks, we find that setting $T_k = |\mathcal{B}|/K$ can be a good starting point to achieve significant improvements towards the new task without hurting the performance of other tasks. We believe the lack of negative impact on existing tasks can be attributed to the correlation between tasks. For example, to purchase a product users are likely to click on it first, thus adding the click task will not hurt the performance of the purchase task.

\subsection{Inference and Serving}

\begin{figure}
    \centering
    \includegraphics[width=\columnwidth]{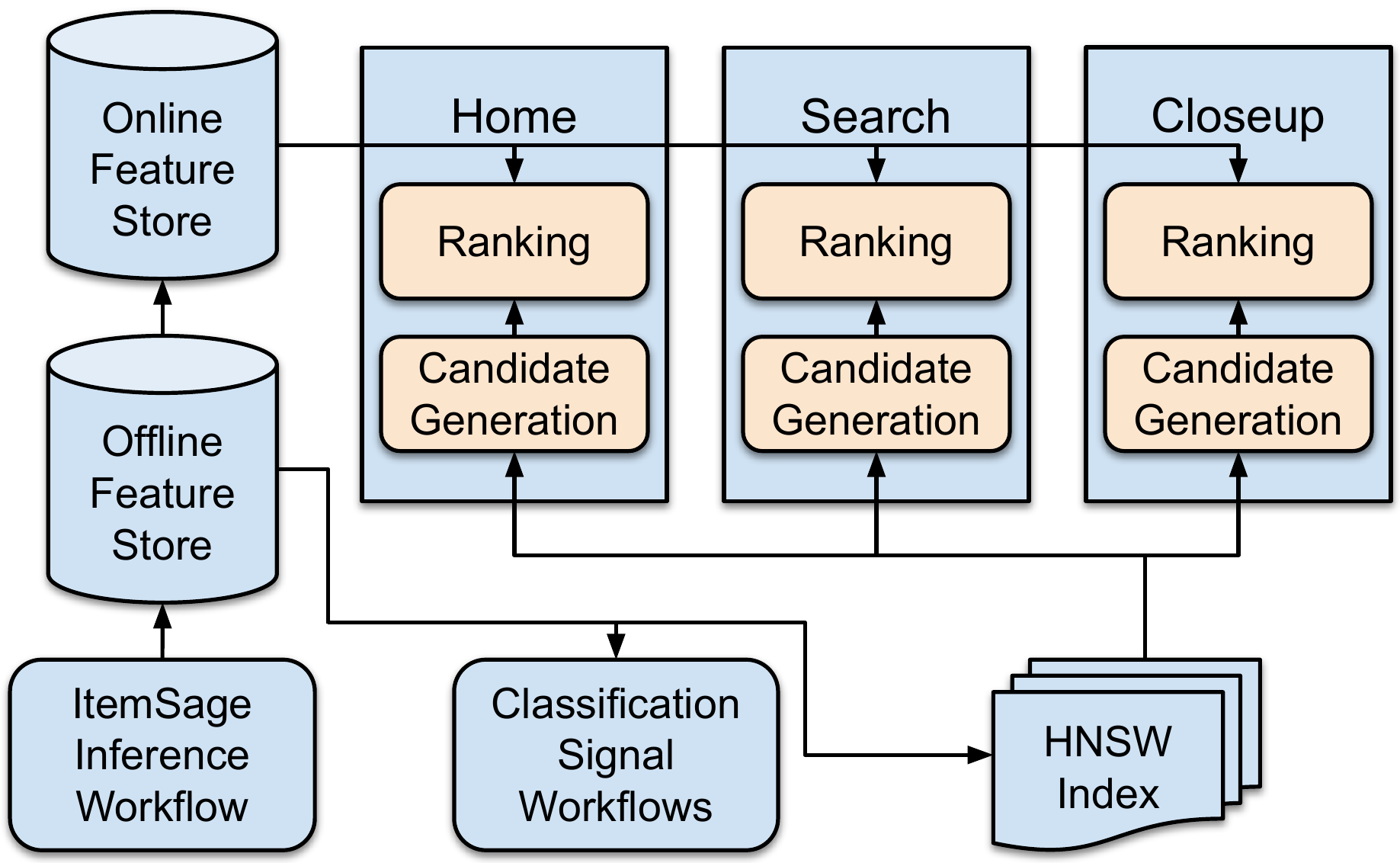}
    \caption{ItemSage inference and serving.}
    \label{fig:serving}
\end{figure}

Figure \ref{fig:serving} illustrates how ItemSage embeddings are deployed to power Pinterest shopping recommendations. The embeddings are inferred in a batch workflow, which ensures that the model inference latency does not impact the end-to-end latency of the vertical recommendation systems. The inference workflow runs daily to update the embeddings based on the latest features and to extend the coverage to newly ingested products. Each new set of embeddings is used to create an HNSW index \cite{hsnw} for candidate generation and is also pushed to the online feature store for ranking applications. The HNSW index and the feature set are reused by all of the different vertical systems for Home, Closeup and Search recommendations. The Home and Closeup surfaces use precomputed PinSage embeddings fetched from the feature store as queries, while in Search, the query embeddings are inferred on the fly to support the long tail of previously unseen queries. The main thing to note is the simplicity of this design enabled by multi-task learning. By creating a single set of embeddings for all 3 vertical applications, we can use a single inference workflow and a single HNSW index to serve recommendations, thereby reducing the infrastructure and maintenance costs by three times.

After ItemSage embeddings are published to the offline feature store, they can be used as features in classification models designed to infer missing information about the products in our catalog. One example with which we have seen early success is inferring whether a product (e.g. in the fashion vertical) is intended to be sold to male users, female users or whether it is unisex. Training a simple MLP on top of ItemSage embeddings yielded a 3.6\% improvement in top-1 accuracy over our previous baseline.\footnote{We do not provide a detailed presentation on using ItemSage for classification applications in this work as our approach does not control for several important factors including dataset, features and model architecture compared to our baseline. Nonetheless, it is encouraging to see that our product embeddings can deliver impact to other applications with little effort.}

\section{Experiments} \label{sec:experiments}

In this section, we provide an empirical analysis of ItemSage embeddings, focused on evaluating the design choices outlined in this paper. We first conduct an extensive evaluation of the embeddings on offline benchmarks, followed by sharing results obtained via A/B experiments on live traffic.

\subsection{Offline Evaluation} \label{sec:offline-eval}

\subsubsection{Evaluation Protocol} \label{sec:eval-protocol}

Our goal is to build embeddings that are effective throughout the recommendation stack starting from candidate generation. Therefore we use recall as the main metric to evaluate the quality of embeddings and potential trade-offs.

We set up eight offline benchmarks for model evaluation, including four engagement types (clicks, saves, add-to-cart actions and checkouts) for two surfaces with different query modalities (Closeup and Search). Each benchmark $\mathcal{P}$ consists of a set of 80,000 pairs of query and engaged products $(x_i, y_i)$ that are sampled from image based recommendations or search results not included in the training data. We also randomly sampled a distractor set $\tilde{\mathcal{C}}$ of one million products from the shopping catalog $\mathcal{C}$ to measure the engagement-weighted recall at $k$, which we define as the weighted proportion of $(x_i, y_i)$ pairs for which the engaged product $y_i$ is ranked within top $k$ among $\tilde{\mathcal{C}} \cup \{y_i\}$,

\begin{equation}
Recall@k = \frac{1}{\sum_{i} w_i} \sum_{i=1}^{|\mathcal{P}|} w_i \mathbf{1}\left\{ \left| \left\{\langle q_{x_i}, p_y \rangle \ge \langle q_{x_i}, p_{y_i} \rangle | y \in \tilde{\mathcal{C}}\right\}\right| \leq k \right\},
\end{equation}
where $w_i$ represents the engagement count associated with $(x_i, y_i)$. In the following experiments, we fix the value of $k$ to 10 since the results for other values of $k$ are similar.

\subsubsection{Model Architecture} \label{sec:arch-experiments}

\begin{table*}[t]
  \caption{Comparison of different model architectures with baselines.}
  \label{tab:architectures}
  \begin{tabular}{l|c|cccc|cccc}
    \toprule
    & Number of & \multicolumn{4}{c|}{Closeup} & \multicolumn{4}{c}{Search}\\
    & Parameters & Clicks & Saves & Add-to-Cart & Checkouts & Clicks & Saves & Add-to-Cart & Checkouts\\
    \midrule
    Sum & - &    0.663 &    0.647 &    0.669 &    0.699 &    - &    - &    - &    - \\
    Sum-MLP & - & - & - & - & - & 0.577 & 0.533 & 0.561 & 0.629 \\
    MLP-Concat-MLP & 30.8M & 0.805 & 0.794 & 0.896 & \textbf{0.916} & 0.723 & 0.736 & 0.834 & 0.861 \\
    ItemSage & 33.1M & \textbf{0.816} & 0.812 & \textbf{0.897} & \textbf{0.916} & 0.749 & 0.762 & \textbf{0.842} & \textbf{0.869} \\
    2-Layer Transformer & 36.3M &    0.815 &    0.809 &    0.895 &    0.913 &    0.745 &    0.759 &    0.837 &    0.867 \\
    3-Layer Transformer & 39.4M &    0.815 &    0.810 &    0.896 &    0.915 &    0.747 &    0.758 &    0.841 &    \textbf{0.869} \\
    4-Layer Transformer & 42.6M &    \textbf{0.816} &    \textbf{0.813} &    \textbf{0.897} &    0.915 &    \textbf{0.750} &    \textbf{0.764} &    0.840 &    \textbf{0.869} \\
    \bottomrule
  \end{tabular}
\end{table*}

In this section we compare the model architecture described above with several baselines and deeper versions of the ItemSage model with additional transformer encoder blocks. The first baseline simply applies sum-pooling and $L_2$ normalization to the PinSage embeddings of the images associated with each product (denoted as Sum). While this baseline is fairly simplistic, note that PinSage embeddings have been independently trained to effectively generate image (pin) candidates for non-shopping use cases at Pinterest. Additionally, other works \cite{amazonsearch} have reported simple pooling as being a baseline difficult to outperform for e-commerce search. 

The Sum baseline cannot generate candidates for search results since the PinSage image model and the fine-tuned DistilBERT models produce embeddings that are not compatible with one another. We introduce a second baseline for search, Sum-MLP, which applies the pretrained MLP for image candidates from SearchSage (introduced in Section \ref{sec:xsage}) on every product image, followed by a sum-pooling and $L_2$ normalization to obtain product level embeddings. 

The Sum and Sum-MLP baselines require little effort to generate. Consequently, they were adopted in production to support shopping recommendations while ItemSage embeddings were under development. We will refer to these baselines again in Section \ref{sec:online-experiments} when discussing results in A/B experiments. 

The final baseline (denoted as MLP-Concat-MLP) is more competitive. It first maps each input feature into a latent space by applying a 3-layer MLP module with 256 hidden units. We learn separate MLPs for image and text features. The latent representations are concatenated into a single vector and passed through a second 3-layer MLP.

The results are presented in Table \ref{tab:architectures}. We observe that ItemSage strongly outperforms the Sum and Sum-MLP baselines. The transformer architecture yields improvements over the MLP-Concat-MLP model baseline on all tasks; the most notable improvements can be seen in the search tasks for clicks and saves. We attribute these gains to the self-attention module in ItemSage. However, using deeper architectures does not significantly improve the model performance: the results using 2 or 3 layers are worse than the 1 layer baseline, while the model with 4 layers has mixed results. In all cases, the relative change in metrics is small and considering the increased cost of deeper architectures, we chose to deploy the variant with a single transformer layer.

\subsubsection{Feature Ablation} \label{sec:feature-ablation}

In this section, we analyze the benefit of learning the ItemSage embeddings from multi-modal features. We compare our final model with two different models using the same transformer architecture, but limited to using features corresponding to a single modality: image or text (Table \ref{tab:ablation}, Row ``Feature''). The image-only model takes as input just the PinSage embeddings of the product's images. The text-only model is trained based on the text attributes from the shopping catalog (title, description, etc.). We observe that the model trained on features from both modalities has significantly better performance than both baselines on all 8 tasks. Also, the image only model has significantly stronger performance over the text-only model, which could be attributed to the additional information summarized into the image embeddings: (1) PinSage is a graph neural network (GNN) aggregating information from the Pinterest pin-board graph in addition to the image itself and (2) the learnings presented in this paper regarding multi-modal learning have also been applied within PinSage to textual metadata available with each image. As one might expect, the models trained on a single feature modality have stronger performance when the query comes from the same modality, i.e., the image-only model shows better performance on Closeup tasks, while the text-only model performs better on Search tasks.

Inspired by PinSage and related work on GNNs \cite{graphsage, gtn1}, we conducted an experiment augmenting ItemSage with information obtained from the Pinterest pin-board graph \cite{pixie}. Products can naturally be embedded into this graph by creating edges between a product and its own images. For each product, we performed random walks starting from its corresponding node (reusing the same configuration as PinSage \cite{pinsage}) and kept the most frequently occurring 50 image neighbors that are different from the product's own images. The images are mapped to their corresponding PinSage embeddings which are appended to the sequence of features provided as input to the transformer. This extension to ItemSage showed neutral results (Table \ref{tab:ablation}, Row ``Feature'') and increased training cost. We attribute this result to the fact that the PinSage embeddings of the product's own images are already aggregating the graph information, making the explicit extension redundant.

\subsubsection{Loss Ablation} \label{sec:loss-ablation}

In Section \ref{sec:loss}, we introduce our approach for sampling negative labels which consists of two sources: (1) other positive labels from the same batch as the current example and (2) randomly sampled negative labels from the entire catalog. In this section, we compare how our model performs if the negative labels are selected from only one of these sources. The results are presented in the Row ``Negative Sampling'' of Table \ref{tab:ablation}. We observe a steep drop in recall if only one source of negatives is used. Moreover, if we only select random negatives, the model converges to a degenerate solution (and thus we need to apply early stopping) where a few products become very popular and appear in the top 10 results of more than 10\%-15\% of the queries in the evaluation set depending on the task. 

We also compare our mixed negative sampling approach with the approach from \cite{yang2020mixed}, and observe that our approach which introduces separate loss terms for in batch positives and random negatives provides an improvement of at least 3.5\% on every task.

\subsubsection{Task Ablation} \label{sec:task-ablation}

\begin{table*}[t]
  \caption{Ablation study for ItemSage models. Relative differences from the ItemSage model are shown in the parentheses.}
  \label{tab:ablation}
  \begin{tabular}{ll|cccc|cccc}
    \toprule
    && \multicolumn{4}{c|}{Closeup} & \multicolumn{4}{c}{Search}\\
    && Clicks & Saves & Add Cart & Checkouts & Clicks & Saves & Add Cart & Checkouts\\
    \midrule
    \multicolumn{2}{c|}{ItemSage} 
        & 0.816 &    0.812 &    0.897 &    0.916 &    0.749 &    0.762 &    0.842 &    0.869 \\
    \midrule
    \multirow{6}{40pt}{Feature} 
    & \multirow{2}{*}{Image Only}
     &    0.795 &    0.787 &    0.882 &    0.908 &    0.670 &    0.698 &    0.798 &    0.830 \\
    && (-2.6\%) & (-3.1\%) & (-1.7\%) & (-0.9\%) &(-10.5\%) & (-8.4\%) & (-5.2\%) & (-4.5\%) \\
    & \multirow{2}{*}{Text Only}    
     &    0.683 &    0.658 &    0.832 &    0.859 &    0.669 &    0.665 &    0.790 &    0.820 \\
    &&(-16.3\%) &(-19.0\%) & (-7.2\%) & (-6.2\%) &(-10.7\%) &(-12.7\%) & (-6.2\%) & (-5.6\%) \\
    & \multirow{2}{*}{Image + Text + Graph}    
     &    0.814 &    0.812 &    0.893 &    0.905 &    0.743 &    0.767 &    0.842 &    0.860 \\
    && (-0.2\%) &  (0.0\%) & (-0.4\%) & (-1.2\%) & (-0.8\%) &  (0.7\%) &  (0.0\%) & (-1.0\%) \\
    \midrule
    \multirow{6}{40pt}{Negative Sampling} 
    & \multirow{2}{*}{$L_{S_{pos}}$ Only}        
     &    0.597	&    0.602 &    0.717 &    0.772 &    0.553	&    0.544 &    0.662 &    0.724 \\
    && (--26.8\%) & (-25.9\%) &(-20.1\%) & (-15.7\%) & (-26.2\%) & (-28.6\%) &(-21.4\%) &(-16.7\%) \\
    & \multirow{2}{*}{$L_{S_{neg}}$ Only}
     &    0.774 &    0.768 &    0.868 &    0.897 &    0.655 &    0.670 &    0.804 &    0.840 \\
    &&(-5.1\%) &(-5.2\%) & (-3.2\%) & (-2.1\%) &(-12.6\%) &(-12.1\%) & (-4.5\%) & (-3.3\%) \\
    & \multirow{2}{*}{$L_{S_{mixed}}$}
     &    0.781 &    0.774 &    0.860 &    0.884 &    0.687 &    0.706 &    0.809 &    0.838 \\
    &&(-4.3\%) &(-4.7\%) & (-4.1\%) & (-3.5\%) &(-8.3\%) &(-7.3\%) & (-3.9\%) & (-3.6\%) \\
    \midrule
    \multirow{4}{40pt}{Surface} 
    & \multirow{2}{*}{Closeup}
     &    0.815 &    0.811 &    0.891 &    0.909 &        - &        - &        - &        - \\
    && (-0.1\%) & (-0.1\%) & (-0.7\%) & (-0.8\%) &        - &        - &        - &        - \\
    & \multirow{2}{*}{Search}
     &        - &        - &        - &        - &    0.760 &    0.766 &    0.830 &    0.861 \\
    && -        & -        & -        & -        & ( 1.5\%) & ( 0.5\%) & (-1.4\%) & (-0.9\%) \\
    \midrule
    \multirow{4}{40pt}{Engagement Type} 
    & \multirow{2}{*}{Clicks + Saves}        
     &    0.819	&    0.812 &    0.869 &    0.894 &    0.755	&    0.768 &   0.689 &    0.765 \\
    && ( 0.4\%) & ( 0.0\%) & (-3.1\%) & (-2.4\%) & ( 0.8\%) & ( 0.8\%) &(-18.2\%) &(-12.0\%) \\
    & \multirow{2}{*}{Add Cart + Checkouts}
     &    0.503 &    0.503 &    0.850 &    0.882 &    0.382 &    0.392 &    0.768 &    0.793 \\
    &&(-38.4\%) &(-38.1\%) & (-5.2\%) & (-3.7\%) &(-49.0\%) &(-48.6\%) & (-8.8\%) & (-8.7\%) \\
    \bottomrule
    
  \end{tabular}
\end{table*}

In this section, we evaluate the effectiveness of the multi-task learning regime used to train ItemSage.

We first evaluate the recall of ItemSage embeddings against two baseline models trained on vertical specific tasks: ``Closeup'' and ``Search'' (Table \ref{tab:ablation}, Row ``Surface''). Each baseline uses the same model architecture and represents the performance we would expect if we deployed separate embeddings per vertical. We find it encouraging that the model optimized for both verticals performs better on 6 out of 8 tasks. This suggests that applying multi-task learning leads to improved performance even across product verticals and, in the cases where the performance degrades, the degradation is not substantial compared to a vertical specific model. This is a remarkable result considering that the PinSage and SearchSage embeddings are not compatible with one another. We also find it interesting that the largest improvements are seen on shopping specific objectives (add-to-cart actions and checkouts), these actions being 30 times more sparse than clicks and saves, suggesting that the cross vertical setup helps the model extract more information about which products are more likely to be purchased by users.

The next experiment focuses on the impact of mixing regular engagement tasks (clicks and saves) together with shopping engagement tasks (add-to-cart actions and checkouts). The results are summarized in the Row ``Engagement Type'' of Table \ref{tab:ablation}. As expected, regardless of the task, we see an improvement in recall whenever we optimize the model on that specific task. Furthermore, we observe substantial gains in add-to-cart actions and checkouts compared to a model specialized at modeling just shopping engagement and minimal losses on clicks and saves tasks compared to the corresponding specialized model. We believe the substantial gains (3.7\%-8.7\%) of the joint model on add-to-cart actions and checkouts can be explained by the significant difference in training data volume between shopping and regular engagement, the additional click and save labels help the embeddings converge towards a more robust representation.

\subsection{Online Experiments} \label{sec:online-experiments}

\begin{table}
  \caption{Results of A/B experiments. The three columns show the relative difference between ItemSage and the baselines in number of total clicks, average checkouts per user, and average Gross Merchandise Value (GMV) per user.}
  \label{tab:online-results}
  \begin{tabular}{c c c c}
    \toprule
    Surface & Clicks & Purchases & GMV \\
    \midrule
Home          & 11.34\% & 2.61\% & 2.94\% \\
Closeup   	  & 9.97\% & 5.13\% & 6.85\% \\
Search        & 2.3\% & 1.5\% & 3.7\% \\
    \bottomrule
  \end{tabular}
\end{table}

We report results from A/B experiments applying ItemSage in each of the main surfaces at Pinterest: Home, Closeup and Search. In these experiments, ItemSage was used to generate candidates for upstream recommendation systems via approximate nearest neighbor retrieval \cite{hsnw}. As shown in Section \ref{sec:task-ablation}, ItemSage is directly applicable to Search and Closeup recommendations; to extend to Home recommendations, we cluster the user activity history and sample pins from several clusters to reduce the problem to the same setting as producing Closeup recommendations \cite{pinnersage}. The embeddings used in the control group are generated with the baselines introduced in Section \ref{sec:arch-experiments}: the Sum baseline is used for Home and Closeup recommendations and the Sum-MLP baseline is used in Search. The results are summarized in Table \ref{tab:online-results}. We observed a strong impact on both engagement and shopping specific key business indicators deploying ItemSage embeddings for product recommendations on all surfaces.

\section{Conclusion}

In this work, we presented our end-to-end approach of learning ItemSage, the product embeddings for shopping recommendations at Pinterest. 

In contrast to other work focused on representation learning for e-commerce applications, our embeddings are able to extract information from both text and image features. Visual features are particularly effective for platforms with a strong visual component like Pinterest, while modeling text features leads to improved relevance, especially in search results. 

Furthermore, we describe a procedure to make our embeddings compatible with the embeddings of other entities in the Pinterest ecosystem \textbf{at the same time}. Our approach enables us to deploy a single embedding version to power applications with different inputs: sequences of pins (images) for user activity based recommendations in the Home surface, single images for image based recommendations in the Closeup surface and text queries to power search results. This leads to a 3X reduction in infrastructure costs as we do not need to infer separate embeddings per vertical. From our experience, embedding version upgrades are a slow process taking consumers up to a year to completely adopt a new version before an old one can be fully deprecated. A unified embedding for all applications means less maintenance throughout this period and a more efficient, consolidated process for upgrades and deprecation. 

Finally, we show that by applying multi-task learning, ItemSage embeddings can optimize for multiple engagement types leading to improved performance for objectives with sparse labels and no penalty for objectives with sufficient training labels. By optimizing our product embeddings this way, we can ensure that the shopping recommendation stack is efficient with respect to all objectives starting from the candidate generation layer. 

The effectiveness of our approach is demonstrated by thorough offline ablation studies and online A/B experiments. There are several promising areas for future work, such as replacing the bag of words model used for text features with a pretrained Transformer model or using neighborhood sampling to extract additional multi-modal features from the Pinterest entity graph from shopping specific entities (e.g. merchants) or edges (e.g. co-purchases). 

%
\begin{acks}
The authors would like to thank Pak-Ming Cheung, Van Lam, Jiajing Xu, Cosmin Negruseri, Abhishek Tayal, Somnath Banerjee, Vijai Mohan and Kurchi Subhra Hazra who contributed or supported us throughout this project.
\end{acks}

\bibliographystyle{ACM-Reference-Format}
\bibliography{sample-base}

\end{document}